\documentclass[fleqn,usenatbib]{mnras}
\usepackage{newtxtext,newtxmath}
\usepackage[T1]{fontenc}
\DeclareRobustCommand{\VAN}[3]{#2}
\let\VANthebibliography\thebibliography
\def\thebibliography{\DeclareRobustCommand{\VAN}[3]{##3}\VANthebibliography}

\usepackage{enumitem}
\usepackage{graphicx}	
\usepackage{amsmath}	
\defcitealias{Tomida_2013}{T+13}

\title[Recombination in TDEs]{
The impact of recombination during tidal disruption events}
\author[S. Pacuraru et al.]{
Simona Pacuraru,$^{1}$\thanks{E-mail: sxp1082@student.bham.ac.uk}
Clément Bonnerot$^{1}$ and Martin E. Pessah$^{2}$
\\
$^{1}$School of Physics and Astronomy \& Institute for Gravitational Wave Astronomy, University of Birmingham, Birmingham, B15 2TT, UK\\
$^{2}$ Niels Bohr International Academy, Niels Bohr Institute, Blegdamsvej 17, DK-2100 Copenhagen Ø, Denmark \\
}

\date{Accepted 2026 October 16. Received 2026 October 7; in original form 2026 August 18}

\pubyear{\the\year{}}
\newcommand{\SP}[1]{\textcolor{black}{#1}}
\begin{document}
\label{firstpage}
\pagerange{\pageref{firstpage}--\pageref{lastpage}}
\maketitle

\begin{abstract}
During a tidal disruption event, the resulting debris stream cools down adiabatically due to the tidal stretching. As the temperature drops, the gas is expected to undergo chemical processes, which can release thermal energy into the stream, potentially affecting the subsequent gas evolution. For the first time, we investigate in detail this effect and its dynamical impact on the early-time evolution of the stream by making use of three dimensional hydrodynamic simulations coupled with a realistic equation of state. We find that a few days after disruption, the energy injected by hydrogen recombination and molecular hydrogen formation causes the stream thickness to grow much more rapidly. In the bound debris, this effect stops the stream's confinement by self-gravity before the gas reaches apocentre. As a result, the maximum stream thickness increases by a factor that ranges from a few, for the most bound gas, to a few tens for the near-parabolic gas, reaching $\approx 30 \, R_{\star} $ around the peak of the mass fallback rate. We discuss how this accelerated stream expansion may affect the subsequent evolution of the gas, estimate the luminosity powered by recombination in the unbound debris, and evaluate the potential influence of non-ideal magneto-hydrodynamic effects. By characterizing the thermodynamic and hydrodynamic properties of the stream before its return near pericentre, our results provide physically motivated initial conditions to self-consistently model the later stages of tidal disruption events, offering a promising pathway to unveiling the physical origins of their observed emission. 
\end{abstract}




\section{Introduction}
Stellar gravitational encounters in dense galactic nuclei can occasionally launch a star onto a near-parabolic trajectory toward the central massive black hole. A tidal disruption event (TDE) happens when the pericentre of the star falls within the tidal radius, where the intense tidal pull of the black hole overcomes the stellar self-gravity \citep[][]{Rees1988}. As a result, the star is distorted and stretched into a thin and elongated stream of gas, half of which is ejected on unbound trajectories. The remaining half becomes bound and its fallback and subsequent evolution around the black hole produce a luminous multi-wavelength flare of radiation lasting weeks to years. 

The majority of TDEs are highly luminous in the optical band \citep[e.g.][]{vanVelzen,Hammerstein2023}, where the light curves exhibit a rapid rise to peak luminosities $L\sim 10^{43\text{-}45}\rm \, erg \, s^{-1}$, followed by a smooth power-law decline. Near peak, some TDEs are also detected in X-rays \citep[e.g.][]{Guolo_2024}, which likely originate from an accretion disk \SP{formed via stream circularization}. At later times, the viscous evolution of the accretion disk is known to give rise to an optical/ultraviolet plateau emerging hundreds of days post-peak \citep[][]{vanVelzen,Mummery2024Scalings,Mummery2020}. Yet, while this late time behaviour is well-understood, the physical origin of the early time optical peak luminosity remains a subject of intense debate. A primary obstacle is the immense spatial resolution required to accurately resolve the nozzle shock driven by the strong vertical compression of the stream near pericentre \citep[][]{BonnerotLu2022, andalman_2026, nixon2026}. This prevents current simulations from drawing robust predictions regarding the radiative emission during the subsequent stages of the TDE, including the stream self-intersection and the resulting disk formation. \SP{To mitigate this computational bottleneck, several studies have explored localized setups, for instance by injecting an idealized stream to enhance the resolution at the nozzle \citep[e.g.][]{Huang_2024,Meza25,Huang2025}, or by performing local simulations of the stream self-intersection \citep[][]{Jiang2016,LuBonnerot2019}. However, such approaches rely critically on physically motivated initial conditions, demanding an accurate understanding of the early stream dynamics. In this work, we aim to address this by investigating the impact of recombination on the early time evolution of TDE debris streams.}

Prior to disruption, the star is in hydrostatic equilibrium, with core temperatures of $\sim 10^7\, \rm K$, such that the stellar gas is initially fully ionized. Most theoretical studies of TDEs assume that the debris remains fully ionized throughout the entire evolution. However, as the stream expands due to the tidal stretching, it cools down  adiabatically. Once the gas thermal energy becomes comparable to the ionization energy of a given species, free electrons begin to recombine with ions to form neutral atoms, which may subsequently bind into molecules. \SP{By heating up the stream, the ionization energy released by chemical processes could affect the evolution of the debris. This was first suggested by \cite{Kochanek1994}, who included hydrogen recombination in their model through the Saha equation. This pioneering work served as a direct precursor to recent analytic models that similarly account for recombination \citep[][]{Coughlin2023,andalman_2026}, while also motivating the use of a realistic equation of state (EoS) in hydrodynamic simulations \SP{\citep[][]{Steinberg2024,andalman_2026,Martire2026}}.} The global simulations by \cite{Steinberg2024} were the first to show that the heat injected by recombination causes the stream thickness to increase significantly. \cite{andalman_2026} analysed this effect in more detail by combining a semi-analytic model of the stream evolution around the black hole with one-dimensional hydrodynamic simulations of the nozzle shock. They found that hydrogen recombination and molecular hydrogen ($\rm H_2$) formation increase the maximum stream thickness by a factor $\sim 5$ near the peak of the mass fallback rate. By thickening both the incoming and outgoing streams, this effect enhances the likelihood of self-intersection, thereby raising the minimum black hole spin needed to prevent a collision.

Beyond its dynamical impact on the bound part of the stream, recombination has been proposed to power distinctive observational signatures from the unbound debris. \cite{Kasen2010} predicted that homologous expansion would significantly lower the gas optical depth, allowing the recombination energy to escape and produce a brief optical transient. More recently, \cite{Coughlin2023} challenged this result. Using an analytic model of the debris stream evolution, they showed that self-gravity, which was neglected in \cite{Kasen2010}, prevents the stream from rapidly expanding and maintains it very optically thick. As a result, \cite{Coughlin2023} predicted that the recombination energy remains largely trapped, yielding a transient that is significantly less luminous than originally thought.

In this work, we present the first three dimensional hydrodynamic simulations that investigate in detail the dynamical impact of chemical processes during a TDE. \SP{By employing global simulations, our work provides a self-consistent treatment of the stream hydrodynamics, which improves upon \cite{andalman_2026}'s semi-analytic approach. To include the effect of recombination,} we make use of the EoS \SP{introduced by} \cite{Tomida_2013} (hereafter \citetalias{Tomida_2013}), which accounts for hydrogen and helium reactions. By following the stream evolution up until its first return near pericentre, we characterize the hydrodynamic and thermodynamic properties of the stream at early times, providing realistic initial conditions for accurately studying the later stages of the TDE, including the nozzle shock, the stream self-intersection and the disk formation and evolution. We find that the energy injected by hydrogen recombination and $\rm H_2$ formation causes a rapid transverse expansion of the stream, which increases the maximum thickness of the bound debris by a factor that ranges from a few, for the most bound gas, to a few tens for the least bound. Moreover, as the stream infalls towards the black hole, the gas remains neutral and $\rm H_2$ formation keeps its temperature at $2000\text{-}3000 \, \rm K$. Ultimately, the rapid stream expansion triggered by recombination can affect the dissipative and radiative properties of shocks occurring after the stream returns near pericentre, thereby impacting the late time evolution of TDEs and their observational signatures.

The paper is structured as follows. In Section~\ref{S:Methods} we describe the numerical method and the EoS employed for our simulations. In Section~\ref{S:Results}, we present our results regarding the impact of chemical processes on the gas thermodynamics and its effect on the evolution of the stream width. In Section~\ref{S:Discussion} we discuss the results and implications of this work for future studies of TDEs. We summarize our conclusions in Section~\ref{S:Conclusions}. 
\section{Method}\label{S:Methods}
\begin{figure}
\hspace{-0.3cm}
\includegraphics[width=1.075\columnwidth]{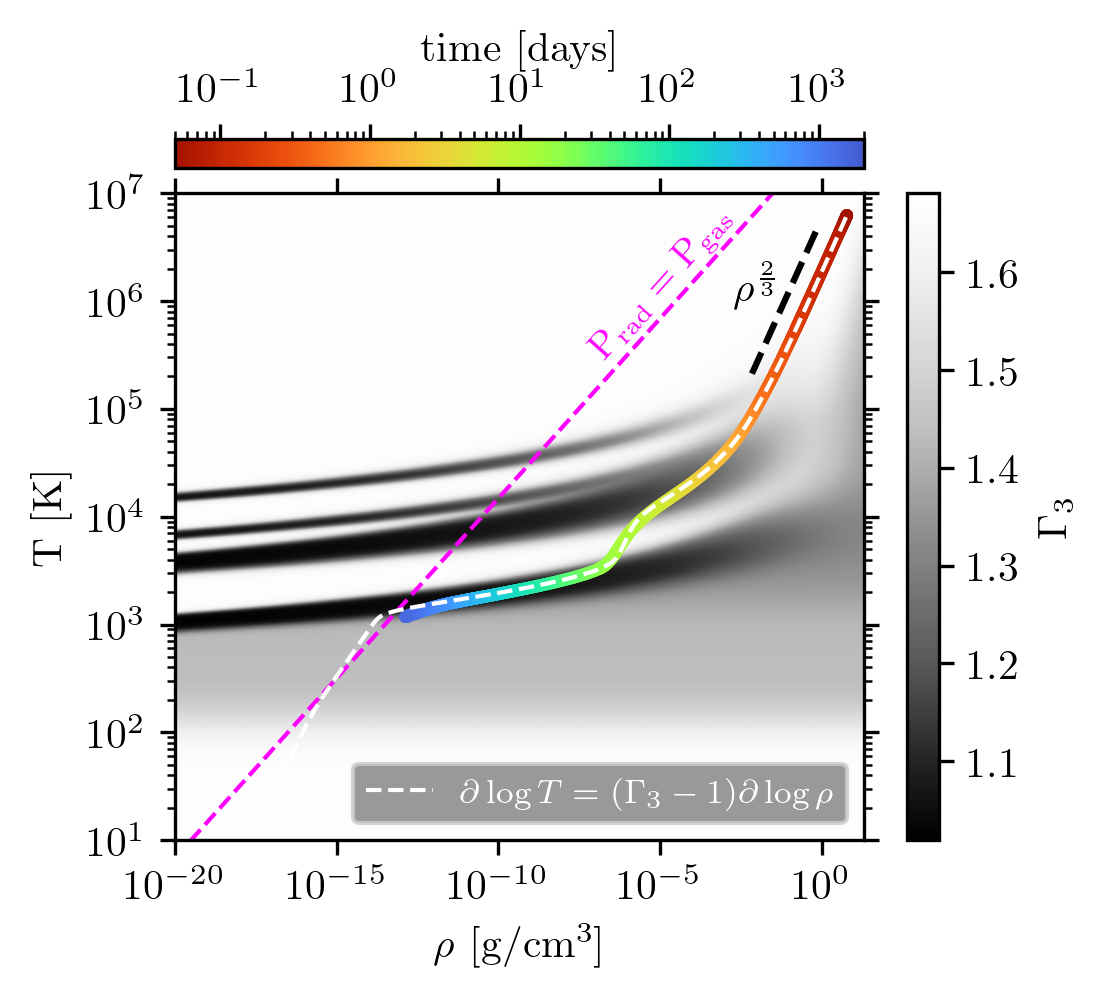}

    \caption{Evolution track of a parabolic section color-coded by time in the $T\text{-}\rho$ phase space color-coded by the third adiabatic index $\Gamma_{3}$. The white dashed line is a solution of $\partial\log{T}=(\Gamma_{3}-1)\partial\log{\rho}$. Above the pink dashed line radiation pressure dominates over gas pressure.}
    \label{fig:T_vs_rho}
\end{figure}

We simulate the tidal disruption of a sun-like star with radius $R_{\star}=R_{\odot}$ and mass $M_{\star}=M_{\odot}$ by a black hole of mass $M_{\rm h}=10^6 \, M_{\odot}$, where $R_{\odot}$ and $M_{\odot}$ are the solar radius and solar mass, respectively. We use \textsc{phantom} \citep[][]{Price2018-phantom} to generate the initial conditions for the star such that it resembles an $n=1.5$ polytrope. \SP{We then evolve the star in isolation using \textsc{gizmo} \citep[][]{Hopkins2015-gizmo} until we obtain an equilibrium configuration. We place the star's centre of mass} on a parabolic trajectory with pericentre equal to the tidal radius $R_{\rm t}=R_{\star}(M_{\rm h}/M_{\rm \star})^{1/3}$, at an initial distance of $3\,R_{\rm t}$ from the black hole, where the influence of the tidal force is negligible. We then simulate the disruption using the Meshless Finite Mass version of \textsc{gizmo}. The gravity solver uses the N-body algorithm inherited from \textsc{gadget} \citep[][]{Springel2005} and we employ an adaptive gravitational softening. In this work, we focus on the evolution of the stream up until its first return near pericentre, therefore we eliminate bound particles that fall below $5\,R_{\rm t}$ on their return trajectory, where the stream becomes underresolved. We set the resolution to $10^6$ particles\SP{, consistently with converged simulations of the initial disruption process and early stream evolution \citep[e.g.][]{Lodato2009,Coughlin2015}.}

To study the impact of chemical processes, we consider two different EoS to evolve the gas: (i) an adiabatic $\gamma$-law  EoS relating pressure $P$, density $\rho$ and specific internal energy $u$ through $P=(\gamma-1)\rho u$, where we set $\gamma=5/3$ constant throughout the simulation, as appropriate for a fully ionized monoatomic gas, and (ii) the more realistic EoS introduced by \citetalias{Tomida_2013}. The latter assumes local thermodynamic equilibrium and uses partition functions to numerically solve the chemical equilibrium equations describing hydrogen and helium recombination/ionization, and $\rm H_2$ formation/dissociation.\footnote{In this work, we only account for hydrogen and helium processes, as metals are not expected to significantly impact the gas thermodynamics. We verified this assumption by using \textsc{mesa}'s \citep[][]{Paxton2011} EoS module to reproduce the map of the adiabatic index shown in Fig.~\ref{fig:T_vs_rho} for solar metallicity, finding no noticeable differences compared to when metals are absent.} The resulting thermodynamic variables are tabulated as a function of ($\log{\rho}$, $\log{T}$), where $T$ is the gas temperature. To implement this EoS in \textsc{gizmo}, we invert the table to express it as a function of ($\log{\rho}$, $\log{u}$), and we linearly  interpolate it to obtain the pressure and the sound speed required to evolve the gas. 
\begin{figure*}
\centering
\includegraphics[width=\textwidth]{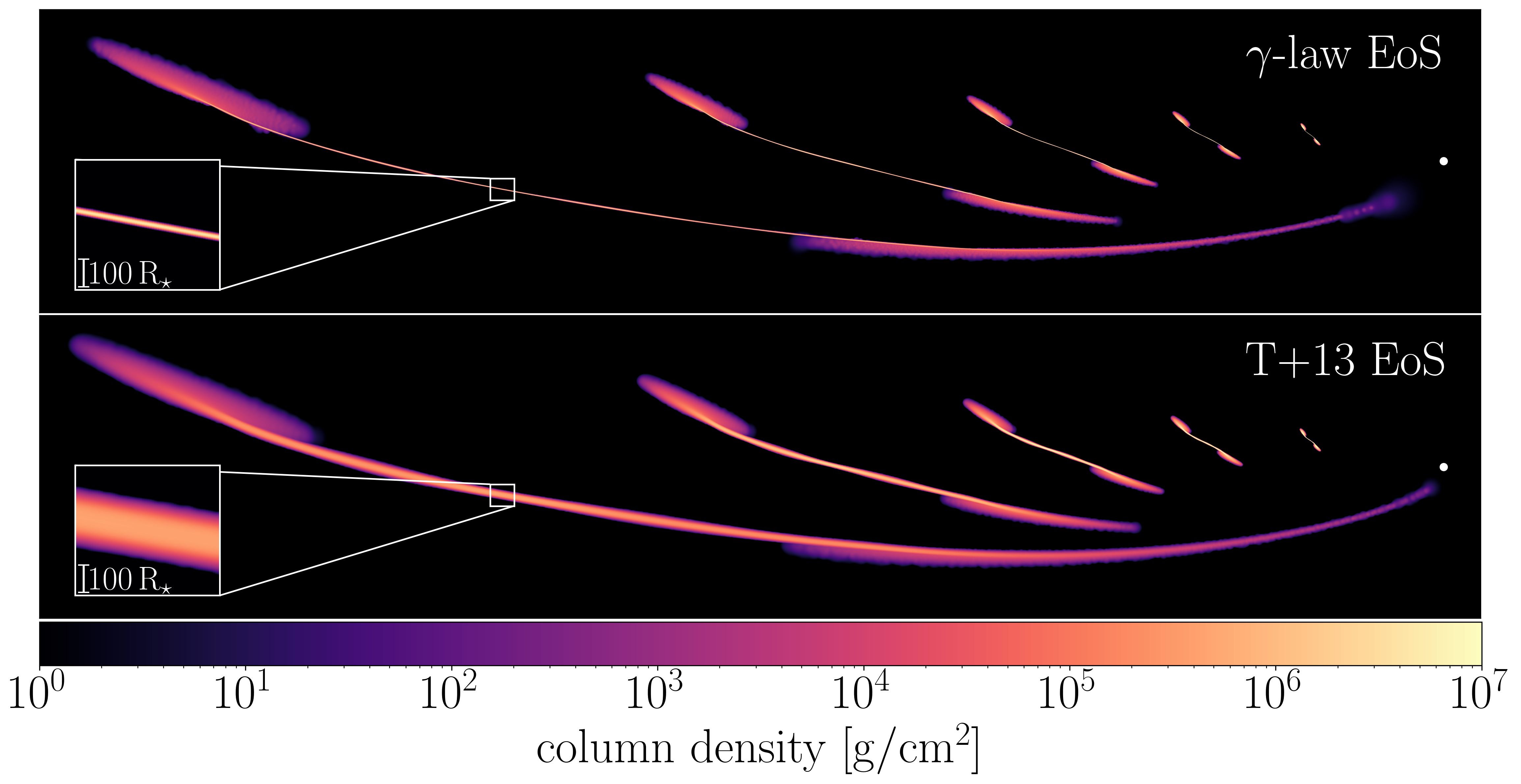}
    \caption{Projection of the density on the orbital plane at time $ t =1.5,3.5,7,14 \text{ and } 26\, \rm days$ after disruption, in the simulation with the $\gamma$-law EoS (top) and with the \citetalias{Tomida_2013} EoS (bottom). In each panel, the snapshots trace the progressive stretching of the debris, from $t=1.5\,\rm days$ after disruption, that is shortly after the stream forms, until $t=26\,\rm days$, where the stream is highly stretched and the bound debris begins to approach the black hole (white dot). The inset panels show a zoom-in of the near-parabolic section of the stream ($|\mu|<0.04$) at $t=26\,\rm days$, when it is located at a distance of $200\, R_{\rm t}$ from the black hole.}
    \label{fig:density_map}
\end{figure*}
With this realistic EoS, the gas thermodynamic evolution can be described by the first and the third generalized adiabatic indices, $\Gamma_{1}\equiv (\partial\log P/\partial\log \rho)_{S}$ and $\Gamma_{3}\equiv 1+(\partial\log T/\partial\log \rho)_{S}$ \citep[see e.g.][]{Chandrasekhar1939} respectively, where $S$ is the gas entropy. For a fully ionized gas $\Gamma_{1}=\Gamma_{3}=5/3$, while for a partially ionized gas both $\Gamma_{1}$ and $\Gamma_{3}$ can approach unity, making the evolution nearly isothermal. This can be seen in Fig.~\ref{fig:T_vs_rho}, showing the $T\text{-} \rho$ phase space color-coded by $\Gamma_{3}$ obtained from the tabulated EoS by \citetalias{Tomida_2013}. In this phase space, the gas evolves along the solutions of $\partial\log T=(\Gamma_{3}-1)\partial\log\rho$. The low-$\Gamma_{3}$ horizontal bands (black regions) correspond, from top to bottom, to the recombination of doubly ionized helium, singly ionized helium, ionized hydrogen and to $\rm H_2$ formation. We note that the \citetalias{Tomida_2013} EoS becomes inaccurate for densities\footnote{For densities $\rho > 0.1 \rm \, g\,cm^{-3}$, non-ideal effects, including particle interactions, pressure ionization of hydrogen and quantum effects, become non-negligible and can alter the gas thermodynamics.} $\rho>0.1\,\rm g\,cm^{-3}$, a condition met initially since the stellar density is $\approx 1\,\rm g\,cm^{-3}$. As we show in Section~\ref{S:Results}, this does not affect our results, as the initial gas evolution in the simulation with the \citetalias{Tomida_2013} EoS resembles that with the $\gamma$-law EoS. Furthermore, the \citetalias{Tomida_2013} EoS doesn't account for the effect of radiation pressure $P_{\rm rad}\approx aT^4/3$, where $a$ is the radiation constant. This becomes significant compared to gas pressure $P_{\rm gas}\approx \rho k_{\rm B} T /m_{\rm p}$ at temperatures $T\gtrsim (3k_{\rm B}\rho/m_{\rm p}a)^{1/3}$ (region above the pink dashed line in Fig.~\ref{fig:T_vs_rho}), where $k_{\rm B}$ is the Boltzmann constant and $m_{\rm p}$ is the proton mass. As we discuss in Section~\ref{S:Discussion}, this only occurs towards the end of our simulation, a few years after disruption.

\section{Results}\label{S:Results}
\subsection{Thermodynamic evolution}\label{S:Thermodynamic evolution}
In Fig.~\ref{fig:T_vs_rho} we show the evolution track of a parabolic section of the stream in the $T\text{-}\rho$ phase space in the simulation with the \citetalias{Tomida_2013} EoS (solid line color-coded by the time in days since disruption). This is obtained by considering a thin slice of the stream with "boundness" parameter $\mu=\epsilon/\Delta\epsilon$ within a narrow range $\pm\Delta\mu$ around $\mu=0$, where $\epsilon$ is the specific orbital energy, $\Delta\epsilon = GM_{\rm h}R_{\star}/R_{\rm t}^2$ is the spread in orbital energy imparted by the disruption under the frozen-in approximation \SP{\citep[e.g.][]{Stone2013}}, and $G$ is the gravitational constant. The temperature and density shown are averaged at different times over the particles belonging to this slice. Initially, the stellar gas is fully ionized, such that following the disruption the temperature decreases as $T \propto \rho^{\Gamma_{3}-1} \propto \rho^{2/3}$. A few hours after disruption, when the temperature falls below $\approx 54.42\,\rm eV/k_{\rm B}=6.3\times 10^5 \, \rm K$, doubly ionized helium starts to recombine into singly ionized helium, and subsequently into neutral helium. Because these processes occur at high densities where the adiabatic index remains relatively large\footnote{\SP{This behavior follows directly from the Saha equation, which predicts that at higher densities the partial ionization zone shifts to higher temperatures. In these conditions, the ionization energy is a smaller fraction of the gas internal energy, diminishing the impact of recombination on the equation of state and causing the low-$\Gamma_{3}$ bands to fade.}} (see fading of the top low-$\Gamma_{3}$ bands in Fig.~\ref{fig:T_vs_rho}), helium has a negligible impact on the gas thermodynamics. The evolution of the parabolic section begins to deviate from the initial power law $T\propto \rho^{2/3}$ roughly one day after disruption, when the temperature drops below $\approx 13.6 \,\rm eV/k_{\rm B}= 1.5\times 10^5 \, \rm K$ and the gas begins to recombine into neutral hydrogen. Unlike helium recombination, the ionization energy released by hydrogen recombination raises significantly the gas thermal energy.\footnote{This assumes that the energy released by recombination is entirely converted into thermal energy, which requires the gas to be optically thick. As discussed in Section~\ref{S:Discussion}, this assumption is valid until a few years after disruption.} This slows down the cooling due to adiabatic expansion and lowers the adiabatic index to $\Gamma_{3} \approx 1.3$, such that the temperature is kept near $\sim 10^4 \, \rm K$. Approximately $10 \, \rm days$ later, when the density drops below $\rho \sim 10^{-6} \rm g\,cm^{-3}$, hydrogen recombination is complete and the gas begins to form $\rm H_2$, which further decreases the adiabatic index to $\Gamma_{3}\approx 1.1$, keeping the temperature around $\sim 10^3 \, \rm K$ hereafter. 

By adopting an $n=1.5$ polytrope, the stellar temperature profile follows $T\propto \rho^{2/3}$, matching the behaviour of a fully ionized gas. As a result, gas travelling on non-parabolic trajectories follows the same $T\text{-}\rho$ evolution track as the parabolic one. This would not be the case for a $n=3$ polytrope (appropriate for a radiative star) which yields a flatter temperature profile $T\propto \rho^{1/3}$, causing the low density regions of the star to be initially hotter compared to the $n=1.5$ case. As a result, for an $n=3$ polytrope we expect hydrogen recombination to last longer in the more dilute parts of the stream. In general, stream elements that are initially more dilute are also cooler and therefore they will undergo chemical processes earlier than denser ones. Unbound and parabolic stream elements will keep expanding indefinitely, whereas bound ones reach their lowest density at apocentre, before compressing and reheating as they infall towards the black hole.

\subsection{Width evolution}\label{S: Width evolution}
As previously found by \cite{Steinberg2024} and \cite{andalman_2026}, our simulations show that the increase in thermal energy caused by hydrogen recombination and $\rm H_2$ formation causes a rapid transverse expansion of the stream. This effect can be seen in Fig.~\ref{fig:density_map}, showing a density map of the stream in the plane of the orbit at different times after disruption. As illustrated by the inset panels at $26\,\rm days$ after disruption, when the \citetalias{Tomida_2013} EoS is used (bottom panel), the whole stream becomes thicker by a factor of a few compared to the fully ionized case (top panel).

\begin{figure}
\centering
\includegraphics[width=\columnwidth]{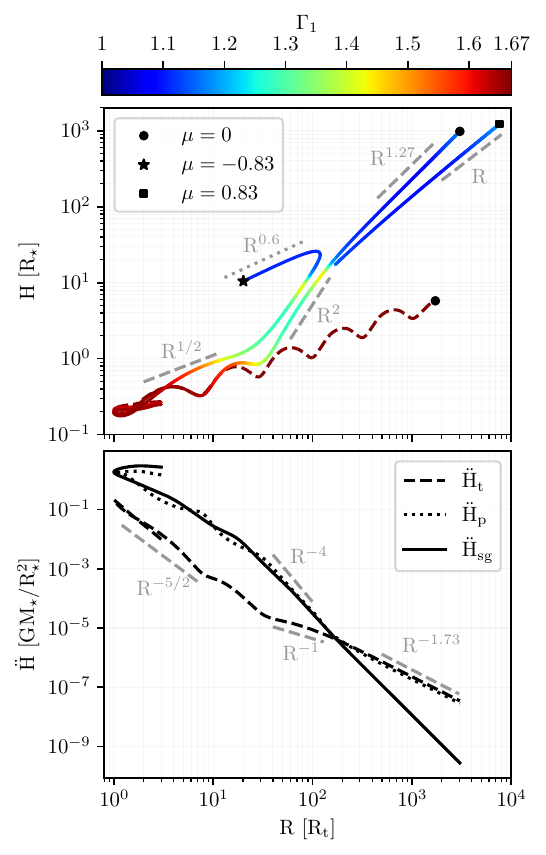}
    \caption{Top: Evolution of the stream width as a function of distance from the black hole for different stream elements of orbital energy $\mu=0$ (dot), $\mu=-0.83$ (star) and $\mu=0.83$ (square), with the \citetalias{Tomida_2013} EoS (solid lines color-coded by $\Gamma_{1}$) and the $\gamma$-law EoS (red dashed line). The grey dashed segments correspond to analytic scalings discussed in Section~\ref{S: Width evolution}, while the grey dotted line is measured in the simulation. Bottom: Evolution of the tidal $\ddot{H}_{\rm t}$, gas pressure $\ddot{H}_{\rm p}$ and self-gravity $\ddot{H}_{\rm sg}$ accelerations for a parabolic section ($\mu=0$). The grey dashed segments are the expected analytic scalings.}
    \label{fig:H_ddotH_vs_R}
\end{figure}

We study in detail the onset and the properties of this expansion effect by measuring the local vertical thickness of the stream $H$.\footnote{We only focus on the vertical width of the stream and not on the in-plane one, because the two are expected to evolve \SP{similarly over the phase of evolution we investigate \citep[][]{Kochanek1994,Bonnerot2022a}.}} This is computed in the simulation by slicing the stream in thin sections selected from a narrow range of boundnesses $\mu \pm \Delta \mu$. The width is then obtained by taking the distances of each particle in the section from its centre of mass and averaging their projections along the vertical direction. We show the resulting width evolution as a function of the distance from the black hole $R$ for different stream sections in the top panel of Fig.~\ref{fig:H_ddotH_vs_R}. In the following, we describe the evolution of the stream width for parabolic, bound and unbound stream elements.

\begin{figure*}
    \centering
    \begin{minipage}[t]{0.48\textwidth}
        \centering
        \includegraphics[width=\linewidth]{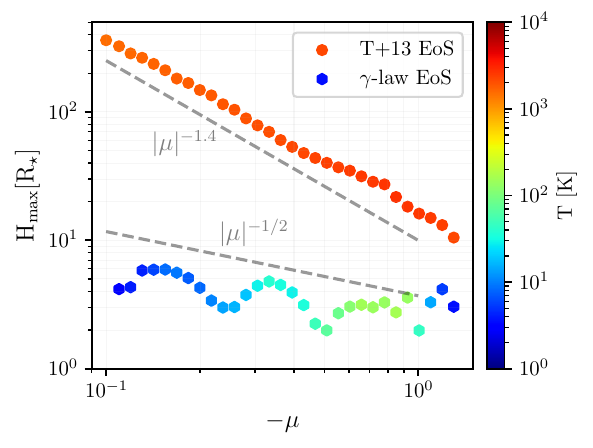}
        \vspace{-0.35cm}
        \caption{Maximum width of bound stream elements color-coded by temperature as a function of boundness $-\mu$ in the simulation with the \citetalias{Tomida_2013} EoS (dots) and with the $\gamma$-law EoS (hexagons). The bottom grey dashed line is the expected analytic scaling for the case with fixed $\gamma$, while the top one is the scaling measured in the simulation with the \citetalias{Tomida_2013} EoS.}
    \label{fig:Hmax_vs_mu}
    \end{minipage}%
    \hfill%
    \begin{minipage}[t]{0.48\textwidth}
        \centering
        \raisebox{0.007cm}{\includegraphics[width=0.919\linewidth]{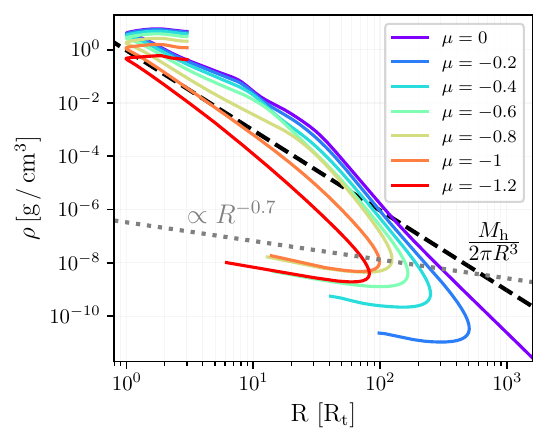}}
        \caption{Density vs distance from the black hole for different stream elements. The black dashed line corresponds to the critical density below which the tidal force dominates over self-gravity. The dotted grey line corresponds to the scaling expected during infall.}
    \label{fig:rho_vs_R}
    \end{minipage}
\end{figure*}
\subsubsection{Parabolic orbit}
The vertical motion of the parabolic section is initially driven by the equilibrium between self-gravity $\ddot{H}_{\rm sg}\approx G\rho H$, and gas pressure acceleration $\ddot{H}_{\rm p}\approx P/(\rho H) \propto \rho^{\Gamma_{1}-1}/H$, where $\rho \propto H^{-2}\ell^{-1}$ is the gas density, and $\ell$ is the local elongation \citep[][]{Bonnerot2022a} which grows as $\ell \propto R^{2}$ at early times. Using these scalings, the hydrostatic balance  $\ddot{H}_{\rm p} \propto \ddot{H}_{\rm sg}$ causes the width to increase as \citep[][]{BonnerotStone2021}
\begin{equation}\label{eq:H_hydrostatic}
    H \propto R^{(2-\Gamma_{1})/(\Gamma_{1} -1)}. 
\end{equation}
For a fully ionized gas $\Gamma_{1} = 5/3$, and equation~(\ref{eq:H_hydrostatic}) therefore gives $H  \propto R^{1/2}$ \citep[][]{Coughlin2016}, which is in good agreement with the width increase found in the simulation with fixed $\gamma$ (red dashed line).\footnote{\SP{The fluctuations found in the stream width evolution (top panel of Fig.~(\ref{fig:H_ddotH_vs_R})) are physical oscillations triggered by the tidal perturbation of the star during its initial passage at pericenter \citep[see e.g.][]{Kochanek1994,Coughlin2016,Bonnerot2022a,andalman_2026}}.}

The resulting width behaviour can be understood by looking at the relative importance of gas pressure $\ddot{H}_{\rm p}$, self-gravity $\ddot{H}_{\rm sg}$ and tidal acceleration $\ddot{H}_{\rm t}$, which are shown for a parabolic section in the bottom panel of Fig.~\ref{fig:H_ddotH_vs_R}. At $R \approx 10 \, R_{\rm t}$, when hydrogen recombination begins, the equilibrium $\ddot{H}_{\rm p} \propto \ddot{H}_{\rm sg}$ is maintained (solid and dotted lines in the bottom panel), however this requires the stream to expand much more rapidly due to the decrease in $\Gamma_{1}$ (see equation~(\ref{eq:H_hydrostatic})). For $\Gamma_{1} = 1.33$, equation~(\ref{eq:H_hydrostatic}) yields $H\propto R^2$, consistent with the width growth found in the simulation for this value of the adiabatic index (see green part of the solid line ending with a dot in the top panel). This fast expansion causes the gas pressure and self-gravity accelerations $\ddot{H}_{\rm p}\propto \ddot{H}_{\rm sg} \propto R^{-4}$ to decrease faster than the tidal one $\ddot{H}_{\rm t} \propto R^{-1}$ (see grey dashed segments in the bottom panel), which becomes significant at $R \approx 100 \, R_{\rm t}$. Hereafter self-gravity no longer affects the stream and the width evolution is driven by an equilibrium between the tidal and gas pressure acceleration $\ddot{H}_{\rm p}\propto \ddot{H}_{\rm t}$ (see dotted and dashed segments in the bottom panel). Using $\ddot{H}_{\rm t}\approx H/R^{3}$ and $\ddot{H}_{\rm p}\propto \rho^{\Gamma_{1}-1}/H$, we find that this equilibrium causes the width to grow as
\begin{equation}\label{eq:H_tidal_pressure_eq}
    H \propto R^{3/(2\Gamma_{1})} \ell ^{(1-\Gamma_{1})/(2\Gamma_{1})}.
\end{equation}
At this stage the gas is undergoing $\rm H_2$ formation, which lowers the adiabatic index to $\Gamma_{1}\approx1.1$. Plugging $\ell \propto R^{2}$ and $\Gamma_{1}=1.1$ into equation~(\ref{eq:H_tidal_pressure_eq}) gives $H\propto R^{\frac{5}{2\Gamma_{1}}-1}\propto R^{1.27}$, which is in good agreement with the late time width growth found in the simulation (see upper panel).

\subsubsection{Unbound orbits}
At early times, the unbound section with $\mu=0.83$ undergoes a fast transverse expansion similar to the parabolic one. The width evolution starts to deviate from the parabolic case once the unbound gas reaches its terminal velocity at a distance $R_{\rm tr}=R_{\rm t}\mu^{-1}(M_{\rm h}/M_{\star})^{1/3}$, where the scaling for the elongation transitions from $\ell \propto R^2$ to $\ell \propto R$. For $\mu=0.83$, this occurs at $R_{\rm tr} = 80 \, R_{\rm t}$. Plugging $\ell \propto R$ into equation~(\ref{eq:H_tidal_pressure_eq}) shows that the equilibrium $\ddot{H}_{\rm p}\propto \ddot{H}_{\rm t}$ should give $H \propto R^{(4-\Gamma_{1})/(2\Gamma_{1})}$. However, for $\Gamma_{1} < 4/3$, this expansion is faster than the free expansion ($H \propto t \propto R$) driven by the gas pressure acceleration alone. Since at this stage the gas is forming $\rm H_2$, $\Gamma_{1}\approx 1.1<4/3$, implying that the equilibrium $\ddot{H}_{\rm p}\propto \ddot{H}_{\rm t}$ cannot be sustained beyond $R_{\rm tr}$. Hereafter, the gas pressure acceleration dominates and the width grows as $H \propto R$ (see solid line ending with a square in the top panel of Fig.~\ref{fig:H_ddotH_vs_R}). Since $R_{\rm tr}$ decreases with increasing $\mu$, less unbound stream elements will take longer to deviate from the parabolic evolution and transition toward homologous expansion.

\subsubsection{Bound orbits}
The width of the bound portion of the stream evolves similarly to the parabolic case as the gas moves away from the black hole, featuring a rapid increase due to the onset of hydrogen recombination. This is shown in Fig.~\ref{fig:H_ddotH_vs_R} for $\mu=-0.83$ (solid line ending with a star marker), corresponding to the gas boundness near the peak of the mass fallback rate. Bound stream elements reach their maximum width near apocentre, before they start infalling toward the black hole. In Fig.~\ref{fig:Hmax_vs_mu} we show the maximum width $H_{\rm max}$ computed along the bound portion of the stream in the simulation, color-coded by the temperature. Using the \citetalias{Tomida_2013} EoS (dot markers) results in a larger $H_{\rm max}$ than in the fixed $\gamma$ case (hexagon markers) everywhere along the stream. In addition, recombination steepens the dependence on $|\mu|$ (see grey dashed lines), causing $H_{\rm max}$ to increase by a factor of a few in the most bound portion of the stream and up to a factor of $\sim 100$ in the least bound. 

This behaviour can be understood by noticing that $H_{\rm max}\approx H(R_{\rm apo})$, where $R_{\rm apo}\approx R_{\rm t}\mu^{-1}(M_{\rm h}/M_{\rm \star})^{1/3}$ is the apocentre distance. In the simulation with fixed $\gamma=5/3$, the width follows $H \propto R^{1/2}$, so that that $H_{\rm max} \propto R^{1/2}_{\rm apo} \propto |\mu|^{-1/2}$ (see bottom grey dashed line). During hydrogen recombination and $\rm H_2$ formation, the width scales between $H \propto R^{1.27}$ and $H \propto R^2$ (see the discussion above for the parabolic section), such that $H_{\rm max} \propto R^\alpha_{\rm apo} \propto |\mu|^{-\alpha}$, where $1.27<\alpha <2$. This explains the steeper scaling $H_{\rm max}\propto |\mu|^{-1.4}$ found in the simulation, where $\alpha\approx 1.4$ is inferred visually from the slope in Fig.~\ref{fig:Hmax_vs_mu} (see top grey dashed line). This also indicates that the width evolution of all bound stream elements can be described by a single mean power law\footnote{This is due to the fact that the tidal force starts dominating around the same distance from the black hole for all stream elements, such that $H\propto R^2$ and $H \propto R^{1.27}$ have the same weight on the overall width dependence, independently of $\mu$.} $H\propto R^{1.4}$. The maximum width can then be approximated as $H_{\rm max} \approx H(R_{\rm rec}) (R_{\rm apo} / R_{\rm rec})^{1.4}$, where $R_{\rm rec}$ is the distance where hydrogen recombination begins. Plugging in $H(R_{\rm rec}) = H_{\star} (R_{\rm rec}/R_{\rm t})^{1/2}$, where \SP{$H_{\star}$ is the width at distance $R_{\rm t}$},\footnote{\SP{$H_{\star}$ depends only on the stellar properties. In particular, the width of a stream section at $R_{\rm t}$ can be approximated as its width within the original star. This is because the tidal force does not distort or compress the star significantly before it reaches pericenter (see top panel of Figure~\ref{fig:H_ddotH_vs_R}).}} gives
\begin{align}\label{eq:H_max}
    H_{\rm max} &= \frac{H_{\star}}{|\mu|^{1.4}}\left(\frac{M_{\rm h}}{M_{\star}}\right)^{0.47} \left(\frac{R_{\rm rec}}{R_{\rm t}}\right)^{-0.9} \notag\\&\approx 400 \, R_{\star} \left(\frac{|\mu|}{0.1}\right)^{-1.4} \left(\frac{M_{\star}}{M_{\odot}}\right)^{-0.47}\left(\frac{H_{\star}}{0.2 \,R_{\star}}\right) \notag\\
    &\phantom{\approx 400 \, R_{\star}} \times \left(\frac{M_{\rm h}}{10^{6}M_{\odot}}\right)^{0.47} \left(\frac{R_{\rm rec}}{10\,R_{\rm t}}\right)^{-0.9},
    \end{align}
which is consistent with the maximum width found in the simulation when recombination is included (see leftmost dot marker in Fig.~\ref{fig:Hmax_vs_mu}). 

In their recent work, \cite{andalman_2026} found that hydrogen recombination and $\rm H_2$ formation increase the maximum width near the peak fallback rate by a factor $\sim 5$. In our set-up, the peak of the fallback rate occurs for $\mu\approx-0.83$, for which we have a factor $\sim 10$ width increase, consistent with their work. As pointed out by \cite{Steinberg2024} and \cite{andalman_2026}, this stream expansion raises the amount of vertical gravitational potential energy that can be converted into kinetic energy during the nozzle shock. Fig.~\ref{fig:Hmax_vs_mu} also shows that, although the gas is much more dilute at the point of maximum width when including recombination, its temperature is significantly higher compared to the $\gamma$-law case. This is because near apocentre the gas is undergoing $\rm H_2$ formation, which keeps the temperature at $\approx 2000-3000\,\rm K$.
    
As found by \cite{andalman_2026}, the transverse tidal acceleration becomes important before the bound stream elements reach apocentre, which can be understood by looking at the gas density evolution. Without recombination, the density scales as $\rho \propto R^{-3}$, remaining above $\rho_{\rm c}\equiv M_{\rm h}/(2\pi R^{3})$ until apocentre, where $\rho_{\rm c}$ is the critical density below which the tidal force dominates over self-gravity. Instead, when recombination is included, the rapid stream expansion causes the density to deviate from this scaling earlier on. This can be seen in Fig.~\ref{fig:rho_vs_R}, showing the density as a function of distance from the black hole for different bound stream elements in the simulation with the \citetalias{Tomida_2013} EoS. Everywhere along the stream, the density drops below $\rho_{\rm c}$ (black dashed line) before reaching apocentre, at which point the tidal force begins to dominate the transverse gas motion. As the stream  infalls towards the black hole, the tidal force causes the width to follow $H \propto R^{0.6}$ (see dotted grey line in Fig.~\ref{fig:H_ddotH_vs_R}), which is similar to the width evolution expected in the fully ionized case $H \propto R^{1/2}$ \citep[][]{Bonnerot2022a}.\footnote{The slightly steeper scaling $H\propto R^{0.6}$ arises because recombination causes the tidal force to dominate before reaching apocentre. As a result, the stream reaches its maximum pre-nozzle width at a point along the orbit of higher curvature than in the fixed $\gamma$ case. This makes the line of intersection between the orbital planes of the fluid elements more inclined relative to the major axis, leading to enhanced compression.} As a result, the thickness of the stream element with $\mu=-0.83$ is expected to reach $\approx R_{\star}$ near pericentre (see solid line ending with a star in Fig.~\ref{fig:H_ddotH_vs_R}).

During infall, the density follows $\rho(R_{\rm t})\propto H^{-2}\ell^{-1} \propto R^{-0.7}$ (see grey dashed line in Fig~\ref{fig:rho_vs_R}), where we have used $H\propto R^{0.6}$ and $\ell \propto R^{-1/2}$. Because of this slow compression, the gas density throughout the stream never exceeds $\sim 10^{-6} \, \rm g\, cm^{-3}$ near pericentre\SP{, keeping the gas temperature at $\approx 2000\text{-}3000\,K$ during fallback.} Since at this density $\rm H_2$ formation is still ongoing (see Fig.~\ref{fig:T_vs_rho}), we expect the gas to be composed by a mixture of hydrogen and $\rm H_2$ prior to the nozzle shock, as also found by \cite{andalman_2026}.

\section{Discussion}\label{S:Discussion}
In this work, we studied with three-dimensional hydrodynamic simulations the impact of chemical processes during the early stages of a TDE. In the following, we discuss the implications of our findings, focusing on the later stages of the TDE, the radiation emitted from the unbound debris, and the potential influence of non-ideal MHD effects.
\subsection{Impact of stream expansion on the later stages of the TDE}
As shown in Fig.~\ref{fig:Hmax_vs_mu}, we find that the thermal energy injected by hydrogen recombination and $\rm H_2$ formation can increase the maximum stream width by a factor of $\sim 10$ near the peak of the mass fallback rate, a result consistent with the works by \cite{Steinberg2024} and \cite{andalman_2026}. Due to this vertical expansion, the compression velocity near pericentre is enhanced, such that the energy available for dissipation at the nozzle shock is increased by two orders of magnitude. \SP{The nozzle shock will heat up the gas significantly, causing it to reionize. As argued by \cite{BonnerotLu2022}, the energy lost in this process is a very small fraction of the thermal energy budget. As a result, reionization is not expected to affect the stream dynamics. As the stream expands past the nozzle shock, a second episode of recombination is triggered. However, since at this point in the evolution the vertical motion is dominated by the tidal force, the thermal energy released by recombination is not expected to alter the ballistic trajectory, except near the next points of maximal tidal compression, where the increased thermal energy can provide a stronger support against collapse \citep[][]{andalman_2026}.}  

Thicker streams \SP{are more} likely to self-intersect despite the precession induced by the Lense-Thirring effect. \SP{This increased likelihood of stream self-intersection holds regardless of the orbital inclination relative to the black hole spin.} Moreover, by lowering the gas optical depth, the stream expansion caused by recombination increases the luminosity produced by the stream-stream collision to $L\approx 6\times 10^{42}(H/30\,R_{\odot})^{2/3}\rm \, erg\,s^{-1}$ \citep[][]{BonnerotStone2021}. Here $H$ represents the stream width at the self-intersection point, which is similar to $H_{\rm max}$ (see Fig.~\ref{fig:Hmax_vs_mu}) provided that the nozzle shock does not significantly affect the transverse evolution following the pericentre passage. The expansion driven by chemical processes can also affect the evolution of the unbound debris as this collides with the ambient medium. As mentioned in \cite{Pacuraru26}, a thicker stream can enhance the synchrotron radiation produced during this interaction, potentially contributing more to the radio emission observed in TDEs \citep[][]{Yalinewich2019}.

\subsection{Recombination-driven emission from the unbound debris}
As originally proposed by \cite{Kasen2010}, and later re-examined by \cite{Coughlin2023}, the energy released by recombination may be radiated away from the unbound debris to produce an optical transient. \cite{Coughlin2023} \SP{assumed} that the recombination energy is insufficient to stop the effect of self-gravity, keeping the stream geometrically thin and optically thick. While our simulations also indicate that the stream maintains hydrostatic balance at the onset of recombination, because of the drop in the adiabatic index, the stream nevertheless undergoes significant transverse expansion (see equation~(\ref{eq:H_hydrostatic}) and the top panel of Fig.~\ref{fig:H_ddotH_vs_R}). This suggests that the gas may eventually become optically thin, supporting the original idea by \cite{Kasen2010}, who assumed homologous expansion $H \propto R$ during the entire evolution. However, while our late time results approach this behaviour (see Fig.\ref{fig:H_ddotH_vs_R}), neglecting the early time hydrostatic phase led \cite{Kasen2010} to underestimate the time of peak luminosity to roughly one week. Moreover, \cite{Kasen2010} underestimated the opacity by several orders of magnitude \citep[][]{andalman_2026}, and thus overestimated the peak luminosity $L$ of the recombination transient, predicting $L\sim 10^{40-42}\rm erg\, s^{-1}$. 

In light of these previous analytic works, we use the results of our hydrodynamic simulations, together with analytic estimates, to infer the radiative properties of the recombining gas. We start by noticing that radiation can only escape once the vertical expansion timescale $t_{\rm exp}=H/\dot{H}$, exceeds the photon diffusion timescale $t_{\rm diff}=\tau H/c = \kappa \rho H^2 /c$, where $\tau$ is the optical depth, $c$ is the speed of light and $\kappa$ is the opacity. Given the high stellar optical depth ($\tau_{\star} \approx \kappa_{\rm es} \rho_{\star} R_{\star} \sim 10^{10}$, where $\kappa_{\rm es}$ is the electron scattering opacity), $t_{\rm diff}$ is initially much larger than $t_{\rm exp}$. However, the disruption and the subsequent stream expansion cause the photon diffusion timescale to quickly drop as $t_{\rm diff} \propto \rho H^2 \propto \ell^{-1} \propto R^{-2}$, while the expansion timescale increases as $t_{\rm exp} \propto R^{3/2}$. By tracking these timescales in the simulation, we find that, for near parabolic trajectories, $t_{\rm diff}$ falls below $t_{\rm exp}$ roughly $10$ days after disruption, when the gas is beginning to undergo $\rm H_2$ formation (see Fig.\ref{fig:T_vs_rho}). We also note that, at this stage, the stream is gas-pressure dominated (see pink dashed line in Fig.~\ref{fig:T_vs_rho}). Consequently, the timescale over which radiative losses cool the gas significantly $t_{\rm cool}\sim (P_{\rm gas}/P_{\rm rad})\,t_{\rm diff}$ is much longer than the photon diffusion timescale, ensuring that the adiabatic assumption holds \SP{\citep[see also][]{andalman_2026}}.

The resulting luminosity can be estimated as
\begin{equation}\label{eq:rec_luminosity}
        L=\frac{\pi e_{\rm rad} H^2 \ell}{t_{\rm diff}}=\frac{\pi a c}{\kappa_{\rm R}} \frac{\ell}{\rho}T^4,
\end{equation}where $e_{\rm rad}=aT^4$ is the radiation energy density and $\kappa_{\rm R}=0.04\, \rm cm^{2}/g$ is the Rosseland mean opacity found by \cite{andalman_2026} when the gas is neutral. The second equality in equation~(\ref{eq:rec_luminosity}) holds because the high optical depth at this stage of the evolution ($\tau \sim 10^4$), together with the local thermodynamic equilibrium assumption, allow us to treat the gas as a black body emitter. Equation~(\ref{eq:rec_luminosity}) shows that, during $\rm H_2$ formation, when the evolution becomes nearly isothermal, the emerging luminosity increases as $L \propto (H \ell )^2$. Plugging into equation~(\ref{eq:rec_luminosity}) the density and temperature computed in our simulation for a parabolic stream section and using $\ell=R_{\star}(R/R_{\rm t})^2$, we find that roughly a year after disruption the luminosity rises to
\begin{equation}\label{eq:luminosity_value}
\begin{aligned}
L \approx 10^{39} \, \mathrm{erg\,s^{-1}} & \left(\frac{\kappa_{\mathrm{R}}}{0.04 \, \mathrm{cm^2\,g^{-1}}}\right)^{-1} \left(\frac{\rho}{10^{-11}\,\mathrm{g\,cm^{-3}}}\right)^{-1} \\
& \times \left(\frac{T}{1800 \, \mathrm{K}}\right)^4 \left(\frac{R}{10^3 \, R_{\mathrm{t}}}\right)^2.
\end{aligned}
\end{equation}
Beyond this time, the optical depth drops below $\sim 10$ and equation~(\ref{eq:rec_luminosity}) becomes less reliable.

The expectation is that the luminosity will continue to increase until the gas becomes optically thin ($\tau\approx1$), which in our simulation occurs a few years after disruption. However, given that $\rm H_2$ formation is nearly complete by this time (see Fig.~\ref{fig:T_vs_rho}), we speculate that the remaining available energy is largely depleted, making the resulting luminosity unlikely to be significant. We note that, a few years after disruption, also the adiabatic assumption breaks down. This is because, when the gas density falls below $\sim 10^{-12} \, \rm g\,cm^{-3}$, radiation pressure $P_{\rm rad}$ becomes comparable to gas pressure $P_{\rm gas}$ (see pink dashed line in Fig.~\ref{fig:T_vs_rho}). As a result, the cooling timescale becomes similar to the photon diffusion timescale and we expect the gas to deviate from the evolution track shown in Fig.~\ref{fig:T_vs_rho}.

\subsection{\SP{Influence of MHD effects}}
While this work advances previous hydrodynamic studies of TDEs by incorporating chemical processes, a more realistic treatment must also account for magnetic fields. In our previous work \citep{Pacuraru26}, we studied the impact of magnetic fields during a TDE, finding that they can become dynamically important and drive a rapid growth of the stream thickness. \SP{This expansion dominates initially over the one caused by recombination if the stellar magnetic field is above $\sim 10^6\, \rm G$. However, by causing the stream to expand earlier, magnetic fields can anticipate the onset of recombination. To determine which of these two effects prevails in the subsequent evolution, we can evaluate the relative importance between gas pressure $P_{\rm gas}\propto \rho^{\Gamma_{1}} \propto H^{-2\Gamma_{1}}\ell^{-\Gamma_{1}}$ and magnetic pressure $P_{\rm mag}=B^2/8\pi$. Assuming magnetic flux conservation and that the magnetic field $B$ aligns with the stretching direction \citep[][]{Guillochon2017,Bonnerot2017}, $P_{\rm mag}\propto B^2\propto H^{-4}$. Soon after magnetic pressure becomes important, the width initially grows as $H \propto R^{3/2}$ \citep{Pacuraru26}. As a result, $P_{\rm mag}\propto R^{-6}$ and $P_{\rm gas}\propto R^{-5\Gamma_{1}}$, where we also used that the elongation follows $\ell \propto R^2$. This implies that $P_{\rm gas}$ can exceed $P_{\rm mag}$ only if $\Gamma_{1}<1.2$, a requirement that is satisfied in our simulations, where recombination lowers the adiabatic index down to $\Gamma_{1}\approx 1.1$ (see top panel of Figure~\ref{fig:H_ddotH_vs_R}). However, as the stream continues to move away from the black hole, the scaling for the width transitions to $H\propto R^{5/4}$. In this regime, gas pressure can become dominant only if $\Gamma_{1}\lesssim 1.1$, a condition that is just marginally satisfied in our simulations. Based on these estimates, we argue that, in the highly magnetized scenarios, determining whether magnetic fields or recombination driven expansion dominate the stream vertical evolution would require simulations that include both recombination and magnetic fields.}

\citet{Pacuraru26} assumed ideal magneto-hydrodynamics (MHD), which is valid only if the gas is fully ionized. As shown in this work, this condition breaks down early on during a TDE. Therefore, when both magnetic fields and chemical processes are taken into account, non-ideal MHD effects should be considered.

In \cite{Pacuraru26}, we showed analytically that Ohmic dissipation is expected to be the first non-ideal MHD effect to develop in the debris stream. Here we compute the ionization fraction $x=n_{\rm e}/n_{\rm n}$ in the simulation using the electron $n_{\rm e}$ and neutral $n_{\rm n}$ number densities provided by the \citetalias{Tomida_2013} EoS and compare it to the critical ionization fraction for Ohmic dissipation, $x^{\rm Ohm}_{\rm crit}=m_{\rm e}c^2\langle \sigma v \rangle_{\rm en}/(4\pi e^2 H \dot{H})$ (see equation (37) of \citet{Pacuraru26}),\footnote{$m_{\rm e}$ is the electron mass, $e$ is the electron charge and $\langle \sigma v \rangle_{\rm en}=8.28\times 10^{-10} (T/\rm K)^{1/2} \rm cm^3\, s^{-1}$ is the electron-neutral momentum transfer rate \citep[][]{Draine1983}.} where $H$ is computed as described in Section~\ref{S: Width evolution}, and $\dot{H}$ is obtained by averaging the gas velocity along the vertical direction. We find that for gas on near-parabolic trajectories, Ohmic dissipation is expected to become important roughly a year after disruption, when the gas has reached a distance $R\sim 10^3\, R_{\rm t}$ from the black hole and the ionization fraction has fallen below $x^{\rm Ohm}_{\rm crit}\sim 10^{-15}$. We also find that this effect can occur for stream elements with $\mu\gtrsim-0.4$, for which $x^{\rm Ohm}_{\rm crit}$ ranges from $\sim 10^{-15}$ to $\sim 10^{-13}$. \SP{In the case of bound stream elements, the minimum ionization fraction is attained at apocenter, where the gas is cooler.} We note that the critical ionization fraction $x^{\rm Ohm}_{\rm crit}$ derived by \citet{Pacuraru26} is inversely proportional to the characteristic length scale over which the magnetic field varies significantly, which was set equal to the stream width $H$ for simplicity. However, in regions of high magnetic field curvature or field reversals, this length scale can be shorter than $H$ (roughly $0.1\,H$ in the simulations by \citealt{Pacuraru26} with an initially toroidal magnetic field), leading to a higher $x^{\rm Ohm}_{\rm crit}$ than what was estimated in equation (37) of \cite{Pacuraru26}. This effect could make Ohmic dissipation important in a larger portion of the bound stream. 

One consequence of entering this non-ideal MHD regime is that the magnetic flux is no longer conserved, an effect that should be taken into account when constructing the initial conditions for simulating the stream-stream collision and the subsequent disk formation and evolution. We also note that the non-conservation of magnetic flux further complicates the already challenging problem \citep[e.g.][]{Tchekhovskoy2014} of accumulating sufficient flux at the black hole horizon to launch a relativistic jet \citep[][]{BlandfordZnajek}. Non-ideal MHD effects can also favour the onset of magnetic reconnection, a physical process that rearranges the magnetic field topology, leading to an efficient conversion of magnetic energy into kinetic energy and thermal energy. Using the Sweet-Parker model \citep[][]{Sweet1958,Parker1957}, we estimate the timescale for magnetic reconnection to become important as $\tau_{\rm rec}=\sqrt{\tau_{\rm diff}\tau_{\rm A}}$, where $\tau_{\rm diff}\approx 4\pi e^2 x L_{0}^2/(m_{\rm e}c^2 \langle \sigma v \rangle_{\rm en})$ is the timescale associated with Ohmic dissipation and $\tau_{\rm A}=\sqrt{4\pi\rho}L_{0}/B$ is the Alfvén time. In these expressions, $B$ is the magnetic field and $L_{0}$ is a characteristic length scale, which we set equal to the stream width $H$. Given that the magnetic field eventually aligns with the stream elongation \citep[][]{Guillochon2017,Bonnerot2017}, and assuming magnetic flux conservation until the onset of Ohmic dissipation, we approximate the magnetic field as $B=B_{\star}(H_{\star}/H)^2$, where $B_{\star}$ is the stellar magnetic field. Plugging this into the expressions above, we find that for a parabolic stream section at distance $R=10^3\, R_{\rm t}$ from the black hole, the reconnection timescale is
\begin{equation}
\begin{aligned}
\tau_{\mathrm{rec}} \approx 1.75 \, \mathrm{yrs} & \left(\frac{T}{1800\, \mathrm{K}}\right)^{-1/4} \left(\frac{H}{200\, \mathrm{R}_{\star}}\right)^{5/2}
\left(\frac{H_{\star}}{0.2\, \mathrm{R}_{\star}}\right)^{-1}\left(\frac{B_{\star}}{10^6\,\mathrm{G}}\right)^{-1/2} \\
& \times\left(\frac{\rho}{10^{-11}\, \mathrm{g\,cm^{-3}}}\right)^{1/4} \left(\frac{x}{10^{-15}}\right)^{1/2},
\end{aligned}
\end{equation}which is much longer than typical TDE timescales even for strongly magnetized stars, suggesting that magnetic reconnection likely plays a negligible role during a TDE.

\cite{andalman_2026} pointed out that metals such as sodium Na, carbon C and potassium K may dominate the free electron population at temperatures near $T \approx 3000\, \rm K$, providing a significant electron fraction $x \sim 10^{-4} (\rho / 10^{-9} \rm g \, cm^{-3} )^{-1/2}$. We note, however, that due to the exponential decay term in the Saha equation, the ionization fraction is highly sensitive to the gas temperature. During $\rm H_2$ formation it goes down to $1000 \, \rm K$ (see Fig.~\ref{fig:T_vs_rho}), at which point equation~(22) in \cite{andalman_2026} predicts a much lower value $x \sim 10^{-10} (\rho / 10^{-13} \rm g \, cm^{-3} )^{-1/2}$. In addition to this, as mentioned above, magnetic field variations on length scales $\ll H$ could significantly increase the critical ionization fraction $x^{\rm Ohm}_{\rm crit}$ compared to the estimate in \cite{Pacuraru26}. We therefore cannot rule out the onset of non-ideal MHD effects during the early evolution of a TDE. A more quantitative assessment of the free-electron population in the stream and of the emergence of non-ideal MHD effects, will require non-ideal MHD simulations that include metals in the EoS. 

\section{Conclusions}\label{S:Conclusions}
We carried out three dimensional hydrodynamic simulations of TDEs, considering a realistic EoS, to investigate the impact of chemical processes during the stream evolution around a supermassive black hole. For this purpose, we implemented the EoS \SP{introduced in} \citetalias{Tomida_2013} in the code \textsc{gizmo}, which we use to simulate the tidal disruption of a main-sequence star by a supermassive black hole of mass $M_{\rm h}=10^6\, M_{\odot}$. \SP{The role of recombination in the evolution of TDE debris streams was recently studied by \cite{andalman_2026} using a semi-analytic model. Our global simulations improve upon their work by providing a more accurate and self-consistent investigation of the impact of chemical processes during the early stream evolution of a TDE. In particular, we expand} upon \cite{andalman_2026}'s work by analysing the role of recombination on both bound and unbound gas, and by providing a scaling relation for the maximum vertical width attained by the bound debris (equation~(\ref{eq:H_max})).

\SP{
Here, we summarize the main conclusions from our work.}

\begin{enumerate}[label=(\roman*), leftmargin=*, align=left]
    \item As previously found by \cite{andalman_2026}, a few days after disruption, hydrogen recombination and $\rm H_2$ formation drive a rapid growth of the stream thickness. We find that this effect is initially required to maintain hydrostatic balance as the adiabatic index drops due to the onset of gas recombination, causing the stream thickness to increase as $H\propto R^2$. This scaling is later modified to $H \propto R^{1.27}$ once the tidal acceleration becomes comparable to the gas pressure acceleration, at which point the influence of self-gravity becomes negligible (see Fig.~\ref{fig:H_ddotH_vs_R}).\\ 
    \item In the bound portion of the debris, hydrogen recombination and $\rm H_{2}$ formation increase the maximum thickness by a factor that ranges from a few to a few tens, reaching approximately $30 \, R_{\star}$ near the peak of the fallback rate (see Fig.~\ref{fig:Hmax_vs_mu}). Due to this fast expansion, self-gravity becomes negligible earlier on compared to the fixed $\gamma$ case.
    During infall, the vertical motion is ballistic and the temperature is maintained around $2000\text{-}3000\, \rm K$ by $\rm H_2$ formation. Prior to the nozzle shock, the gas is composed of neutral hydrogen and $\rm H_{2}$.\\
    \item In the unbound gas, the width evolution transitions from $H\propto R^{1.27}$ to $H \propto R$ as the stream elements reach their terminal velocity. In this part of the stream, this quick expansion can increase the radiation produced by the interaction between the debris and the ambient medium. We also find that the energy released by recombination can be radiated away, producing luminosities of $\sim 10^{39}\, \rm erg\,s^{-1}$. \\
    \item In the absence of metals, gas recombination can trigger non-ideal MHD effects. For gas on parabolic trajectories, we expect Ohmic dissipation to become important roughly a year after disruption when the ionization fraction falls below $x\sim 10^{-15}$, while for non-parabolic stream elements this occurs when $x\sim 10^{-15}-10^{-13}$. We also find that magnetic reconnection develops on very long timescales and therefore it is unlikely to become significant during a TDE. The magnetic field topology and the presence of metals could affect these results and this should be investigated in future work. \\

\end{enumerate}
By including a realistic EoS, the results of this work provide physically informed stream properties for studying self-consistently the later stages of a TDE, including the nozzle shock, the stream self-intersection, the subsequent disk formation and stream-disk interactions. Simulating these processes under physically motivated initial conditions is crucial for understanding the observational signatures of TDEs, particularly the origin of the early time peak luminosity, which to date remains an open question.
\section*{Acknowledgements}
\SP{We thank the anonymous referee for useful comments that helped improve the manuscript. We also thank Aldo Serenelli, Nicholas Stone and Zack Andalman} for helpful discussions. Funded by the European Union (ERC, Unleash-TDEs, project number 101163093). Views and opinions expressed are however those of the author(s) only and do not necessarily reflect those of the European Union or the European Research Council. Neither the European Union nor the granting authority can be held responsible for them. The research leading to this work received funding from the Independent Research Fund Denmark via grant ID 10.46540/3103-00205B. The work presented here is supported by the Carlsberg Foundation, grant CF25-1297. The computations described in this paper were performed using the University of Birmingham's BlueBEAR HPC service, which provides a High Performance Computing service to the University's research community. We acknowledge the use of SPLASH \citep[][]{Price2007} for the visualization of the results. 

\section*{Data Availability}
The data underlying this paper will be shared on reasonable request to the corresponding author. A public version of the GIZMO code is available at \url{http://www.tapir.caltech.edu/~phopkins/Site/GIZMO.html}.



\bibliographystyle{mnras}
\bibliography{bibliography} 






\bsp	
\label{lastpage}
\end{document}